\documentclass{article}
\usepackage{spconf}

\usepackage{graphicx}
\usepackage{algorithm2e}
\usepackage{subcaption} 
\usepackage{booktabs}
\usepackage{float}
\usepackage{siunitx}
\usepackage{multirow}
\usepackage[T1]{fontenc}
\begin{document}
\onecolumn

\title{Hearables: feasibility of recording cardiac rhythms from single ear locations}
\name{ Metin Yarici, Wilhelm Von Rosenberg, Ghena Hammour, Harry Davies, Pierluigi Amadori, Nico Ling, Yiannis Demiris, Danilo P. Mandic}
\address{Department of Electrical and Electronic Engineering, Imperial College London, SW7 2AZ, UK\\
E-mails: \{metin.yarici16,  d.mandic\}@imperial.ac.uk}


\maketitle

\begin{abstract}
Wearable technologies are envisaged to provide critical support to future healthcare systems. Hearables - devices worn in the ear - are of particular interest due to their ability to provide health monitoring in an efficient, reliable and unobtrusive way. Despite the considerable potential of these devices, the ECG signal that can be acquired through a hearable device worn on a single ear is still relatively unexplored. Biophysics modelling of ECG volume conduction was used to establish principles behind the single ear ECG signal, and measurements of cardiac rhythms from 10 subjects were found to be in good correspondence with simulated equivalents. Additionally, the viability of the single ear ECG in real-world environments was determined through one hour duration measurements during a simulated driving task on 5 subjects. Results demonstrated that the single ear ECG resembles the Lead I signal, the most widely used ECG signal in the identification of heart conditions such as myocardial infarction and atrial fibrillation, and was robust against real-world measurement noise, even after prolonged measurements. This study conclusively demonstrates that hearables can enable continuous monitoring of vital signs in an unobtrusive and seamless way, with the potential for reliable identification and management of heart conditions such as myocardial infarction and atrial fibrillation.
\end{abstract}



\section{Introduction}
The use of wearable technologies for monitoring vital signs has become increasingly widespread in the society, both for recreational and medical purposes. Most often, these devices are integrated into wearable garments and accessories, and are concealed and miniaturised for convenience of the user. The most common choices for wearable vital signs monitoring technologies are smart watches and chest straps. Smart watches typically utilise the photoplethysmogram (PPG) to provide continuous monitoring of pulse and respiration, while chest straps may also record the electrocardiogram (ECG). However, PPGs lack the information necessary to understand the functioning of the heart \cite{allen2007photoplethysmography}, while chest-worn devices are not suitable for everyday use due to their obtrusive nature. Consequently, alternative solutions that provide ECG measurements in a convenient and user-friendly manner have attracted significant research interest.\\

One such solution is the 'hearable' device - a wearable that fits in the ear and can serve both as an audio accessory and a platform for health monitoring \cite{goverdovsky2017hearables, yarici2022ear}. For hearables, the stability of the head relative to the vital organs during sitting, sleeping, walking, and eating results in superior monitoring capability in real-world scenarios, relative to devices attached to the limbs \cite{looney2012ear}. For this reason, wearable devices that are placed on the regions of the skin surface surrounding the ear are also gaining popularity. Da He \textit{et al.} \cite{da_ear} and Casson \textit{et al.} \cite{casson_earECG} demonstrated the effectiveness of behind-the-ear and scalp based recordings of ECG, ballistocardiogram (BCG), and PPG for heart rate monitoring, while Celik \textit{et al.} \cite{celik_dry,celik_nano} provided evidence of similar results when measurements were taken from electrodes placed on the ear and neck. Despite the success of these proof-of-concept studies, the difference in ECG potential from the various head locations is not yet well understood.
\\

To that end, our previous work investigated the ECG potential at different head positions from inside a helmet \cite{von2016smart}. An assessment of the performance of the channels under consideration was conducted with regard to the successful application of an R-peak detection algorithm \cite{chanwimalueang2015enabling}. However, through experimentation alone, it is difficult to clarify the effect of the manifold factors that can influence the sensitivity of the ECG sensors \cite{Wil_earECG}. For example, variations in user movements, muscle activity, brain activity, and electrode-skin contact can all adversely affect the quality of the ECG signal \cite{webster1984reducing,thakor1985electrode}, so that the current studies are limited by either a lack empirical evidence or support from theoretical models
\\

\begin{figure}[h!]
    \centering
    \includegraphics{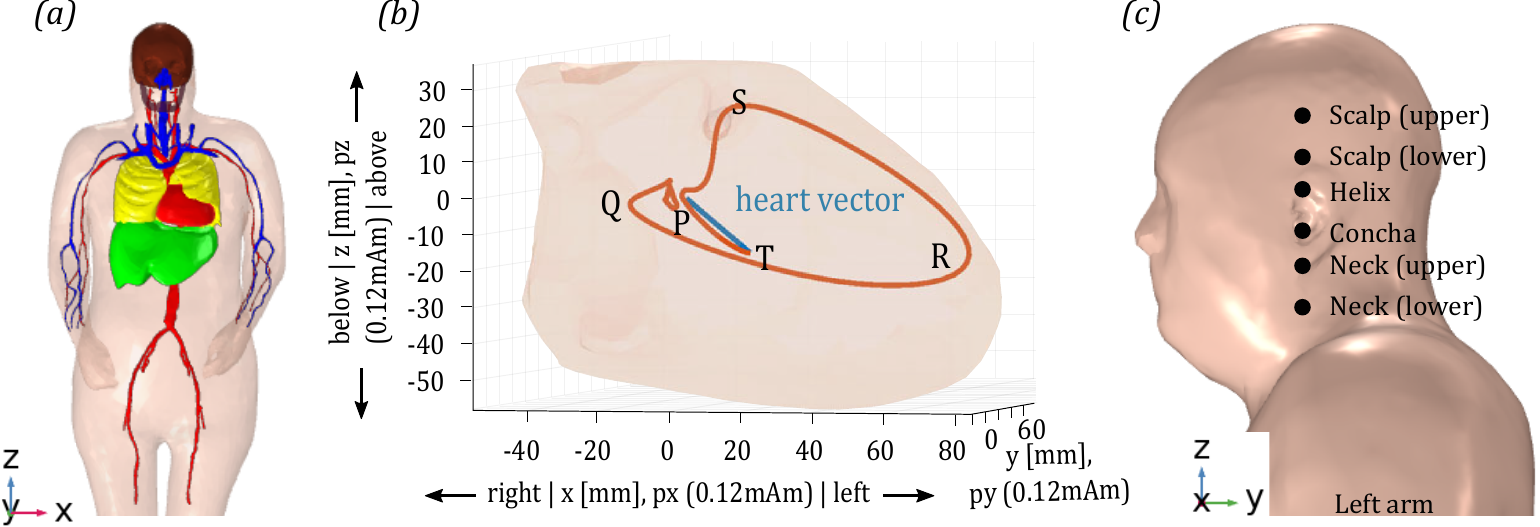}
    \caption{Biophysics model of ECG propagation. (a) 3D Model structure: Organs and tissues in the torso and head are encased in a skin surface structure. (b) Heart vector: The orientation and magnitude of the current dipole p in px, py, and pz; the heart vector at one point in time is shown in blue, and the trace of the tip of the heart vector from the start of the cycle until the current position (axes in 0.12 mAm) is shown in orange; the heart muscle is shown in pink in the background (axes in millimetres). (c) Sensor set-up: For the purpose of mapping the ECG on the left side of the head, sensors were placed on scalp, ear, and neck positions. Potential difference between the upper and lower scalp (scalp ECG), helix and concha (ear ECG), and upper and lower neck electrodes (neck ECG) was extracted. }
    \label{intro_model_sensors}
\end{figure}

A comprehensive theoretical approach to mapping ECG potentials can be achieved through forward modelling \cite{malmivuo1995bioelectromagnetism,gulrajani1998forward}. This process involves applying Maxwell's equations to a dielectric model of the body in order to estimate the propagation of cardiac potentials from the heart to the head-positions of interest. This offers the advantage of examining the ECG in a way that is isolated from other sources of electrical activity, such as brain and muscle activity. Furthermore, such modelling can enable the evaluation of the characteristic ECG - that is, the timings and shape of the waves of the cardiac cyle - available from various wearable ECG channels. Early ECG modelling was based on simplified representations of the human body (for example, a homogeneous conductive sphere \cite{frank1952electric}), however more realistic models have since been produced which include separate shells representing the heart and surrounding body \cite{miller1978simulation}, and realistically shaped geometries for specific organs in the chest cavity \cite{okada2013patient,jiang2009optimization, nakane2019forward, loewe2011determination}. A model which enables the simulation of ECG propagation to the head and ear locations was first introduced in our previous work \cite{Wil_earECG}. A whole-body model incorporating tissue from the torso and head was employed to provide a rigorous theoretical basis for the possibility of high-quality head-ECG. In addition to simulations, a systematic analysis of the ECG surface potential from head and in-ear channels was conducted, measuring the ECG potential difference between the left and right side of the head, as well as between the left and right ears. Moreover, the characteristic timing and shape of the ECG from the different channels was evaluated through both measurements and simulations, opening new avenues for wearable head-based cardiac monitoring that provides more than just heart rate detection. \\

Ideally, wearable vital signs measurements should be conducted from a device worn on a single ear. However, as previously discussed, the ECG potential available on the single ear, or the surrounding locations on the scalp and neck, is not well understood. To this end, we follow the approach outlined in \cite{Wil_earECG} and assess the ECG potential on a single side of the head, ear, and neck. 
\\

This study seeks to determine the amplitude, shape, and timing of ECG waveforms from specific sites through both real-life recordings and biophysics modelling. Once the possibility of recording ECG potentials from each location was established, a second experiment was conducted in a more dynamic, real-world environment, which allowed for the recordings to be evaluated over an extended period of driving in a simulator. This enabled further assessment of the impacts of varying levels of brain and muscle activity, as well as the user's physical movement on the signal quality, and with greater confidence. Results from both experimentation and simulations have established the feasibility of recording ear ECG, and neck ECG signals that possess characteristics similar to those present in the standard Lead I ECG signal. Comprehensive evaluation of performance of the wearable-ECG under consideration relative to Lead I has also been presented. Key findings reveal differences in performance that arise with different recording lengths from each recording channel, thereby revealing the suitability of a given channel for various applications in cardiac monitoring. Overall, our results indicate that the single ear ECG channel and the cross ear ECG channel can provide robust ECG monitoring during prolonged real-world tasks.
\\

Our findings demonstrate the potential of ear ECG for capturing cardiac rhythms with shapes that are very similar to Lead I from the standard limb leads, even during challenging, real-world recordings. This framework holds promise for examining heart conditions that are visible in multiple consecutive cardiac cycles in Lead I, including but not limited to myocardial infarction (the ST segment is elevated), first-degree atrioventricular block (the PR interval is
longer than 200 ms), atrial fibrillation (absence of the P-wave, found in 2\% to 3\% of the population in
Europe and the USA \cite{zoni2014epidemiology}), sinus tachycardia (increased heart rate and a shortened P-T duration) and atrial flutter (an increased frequency of the P-wave relative to the QRS complex as a result of rapid atria contraction) \cite{james2011pocket,hampton2019ecg,lilly2012pathophysiology}. Moreover, the proposed ear ECG framework offers the possibility of 24/7 continuous and unobtrusive cardiac monitoring and can alert the user to the presence of universal signatures of heart malfunction, such as absent P-waves and elongated ST-segments. Such a new perspective can be beneficial to many existing applications, such as remote monitoring for patients with cardiac conditions such as arrhythmias, or heart failure \cite{jahmunah2019computer}, sports and fitness monitoring of athletes \cite{basilico1999cardiovascular}, and assessment of the effect of physical strain and stress in workplace environments \cite{dimsdale2008psychological,knutsson2000shiftwork}. 

\section{Methods}
In the present study, the feasibility of recording cardiac rhythms from a single ear is established through real-world measurements and biophysics simulations. First, theoretical grounding for the measurement of single ear ECG is provided through biophysics modelling of ECG propagation from the heart to various positions of interest around the head and ears, through the use of an accurate volume conductor model of the human body that was first presented in \cite{Wil_earECG}. 

\subsection*{Model geometry}
The geometry within the presented model is based on the VHP-Female Computational Phantom v. 2.1 and v. 2.2 \cite{yanamadala2015new} - a data-set consisting of three-dimensional geometry for major organs and tissues in the human body. The phantom was edited in instances where computational problems were encountered with the mesh of the structures in the model in the modelling software (COMSOL Multiphysics \cite{comsolSoftware}). The model comprised geometric representations for major structures surrounding the heart and in the head, and a complete whole-body volume enveloped by an outer skin structure; the top-half of the model is shown in Figure \ref{intro_model_sensors}a). For the purpose of realistic modelling of electric fields within the body, a large sphere or radius $r=\SI{3.3}{m}$, filled with air, surrounded the body and provided an electrical ground in the model.  The complete mesh consisted of \SI{560,630}{} domain elements and \SI{72,286}{} boundary elements, and the average edge length was \SI{6}{mm}.

\subsection*{Electric properties of the body and cardiac current sources}
 The relevant electric properties of the modelled human tissues (conductivity and relative permittivity at an excitation frequency of \SI{15}{Hz}) were extracted from the database in \cite{hasgall2015database}, and are provided in the supplemental material (Table S1). Previous research has revealed that isotropic and anisotropic treatments of the heart structure in simulations of ECG yield similar results \cite{wei1995comparative}. Therefore, an isotropic treatment of the heart structure was adopted within the present model. Multi-source modelling of the cardiac potential has been shown to perform best for prediction of the ECG potential on the torso \cite{taccardi1963distribution}, however, owing to the relatively large distance between the head and heart, the present model employs single source modelling. The heart source dynamics over the course of a single cardiac cycle are shown in Figure \ref{intro_model_sensors}b; a comprehensive description of the electrical dynamics of the cardiac cycle is provided in \cite{reisner2006physiological}. The source dipole in the presented model is a superposition of three orthogonal vectors representing the projection of the cardiac potential over the course of a single cardiac cycle at a normal to the sagittal, transversal, and coronal planes. The heart vectors were acquired from real-world measurement from a subject with no known cardiac abnormalities. The cardiac cycle was taken to start and end \SI{200}{ms} before and \SI{400}{ms} after the maximum of the R-wave.
 
\subsection*{Experiment A: ECG mapping on the single ear}
ECG potential on candidate positions for head-based wearable ECG monitoring platforms was assessed through measurements on ten subjects with GRASS Ag/AGCl cup electrodes. single ear ECG from the left ear was recorded through one electrode placed on the helix and another on the concha; this reflects two locations that are available to stand-alone, single ear hearable devices. The electrode positioning for the remaining head-ECG channels was devised such that the ear, scalp, and neck channels each spanned equal distances across the head-surface; in this way, differences in ECG potential in each of these regions could be assessed. For the scalp channel, the electrodes were positioned according to the following methodology: the lower scalp electrode was placed at a distance of $L/2$ above the helix-electrode, where $L$ equals the separation between the helix- and concha-electrodes. The upper scalp electrode was placed at a distance of $L$ above the lower scalp electrode. An equivalent placing system was employed for the neck channel, whereby neck electrodes were placed below the concha-electrode. The ear, scalp, and neck electrodes were all positioned along a vertical line which intersected the points at which each subject's helix- and concha-electrodes were positioned (see Figure \ref{intro_model_sensors}c)). A standard limb-ECG channel was also created between the left and right wrist, whereby electrodes were positioned on the left  and right volar central zones (just below the palm) \cite{Wil_earECG, kanna2018bringing}. Prior to the application of the electrodes, the skin at each location was prepared through cleaning with medical wipes and abrasion with NuPrep gel. A layer of 10-20 conductive paste was also applied to the electrodes prior to placement, in order to improve conduction between the skin surface and the electrodes. Measurements were conducted via a custom bio-amplifier programmed to digitise the potential difference between the upper and lower scalp electrodes (scalp ECG), the helix and concha electrodes (single ear ear ECG), the upper and lower scalp electrodes (neck ECG) and the left  and right wrist electrodes (wrist ECG). A sampling rate of \SI{500}{Hz} was used and the helix-electrode served as a bias. During the measurements, subjects were seated and instructed to close their eyes while minimising eye saccades, and movement of the head, neck, and arms. The recordings were performed under the Imperial College London ethics committee approval JRCO 20IC6414, and all subjects gave full informed consent.

\subsection*{Experiment B: real-world feasibility}
Once the cardiac potential on the scalp, ear  and neck  regions had been established for subjects while at rest, the feasibility of recording cardiac rhythms in a real-world scenario was investigated. Single ear measurements on both the left and right ears were conducted, in addition to cross-ear measurements between the left and right ears, for which we previously established the feasibility of recordings on 6 subjects at rest \cite{Wil_earECG,hammour2019hearables, guler2022ear}. Five of the ten subjects from Experiment A were instructed to drive in a virtual reality (VR) driving simulator environment while measurements were taken from the described ear ECG channels and a reference-ECG channel from the wrists. Data was acquired via a g.tec g.USBamp (2011) bio-amplifier at a sampling rate of \SI{1,200}{Hz}. In order to ensure comfortable ear ECG measurements for the duration of the driving task, custom fabric electrodes were used to record the ECG signal from the concha \cite{goverdovsky2015ear}. The fabric electrodes employed within this study were constructed from silver-coated thread that is interwoven with elastic fibres. In \cite{goverdovsky2017hearables}, electrodes of this type were mounted on visco-elastic earpieces and then placed in the ear canal of five subjects over the course of a normal working day. Measurements were shown to be stable after prolonged periods of unrestrained activity which included walking, eating, and talking. While the ear canal is an ideal position to record the ECG, as outlined in \cite{Wil_earECG,hammour2019hearables}, in the current experiment, a PPG sensor was placed in the position of the ear canal for measurement of ear-SpO2, and is the subject of ongoing analysis \cite{davies2020ear}. Therefore, the concha  - another convenient location to record physiological signals from - was used. In the present study, the electrodes were secured to the concha using soft, malleable non-allergenic silicone, shaped to each person's ear. Prior to application of the electrodes, the skin was prepared through cleaning with medical wipes and abrasion with NuPrep gel. A thin layer of Signa Gel conductive gel was applied to the surface of the fabric electrode in order to help establish good conduction at the skin-electrode interface. For the left  and right ear single ear ECG measurements, a reference electrode was positioned on a second location which is suitable for monitoring through Hearable devices - the helix. As with the previous electrode positions, the ipsi-lateral helix was prepared with medical wipes and NuPrep gel prior to electrode connection. The GRASS Ag/AgCl electrodes were used to record the signal from the helix. A layer of 10-20 conductive paste was applied to aid conduction at the skin-electrode interface. As a bias electrode, a second Ag/AgCl electrode was placed on the ipsi-lateral earlobe, with the same skin preparation. For the cross-ear ECG measurement, the left concha electrode was referenced to the right helix electrode, with the left ear lobe electrode serving as the bias. For the reference ECG measurement, for Subjects 1-3, a wrist ECG measurement was conducted, whereby the left wrist electrode was referenced to the right wrist electrode. For Subjects 4 and 5, a reference ECG signal was obtained by referencing the left wrist electrode to the right helix electrode. All reference channels were biased with the left ear lobe electrode. Vertical electro-oculogram (VEOG) channels were also recorded via one electrode below and another above each eye. The previously described skin preparation and conductive paste was also applied for the VEOG measurements.

\subsection{Signal processing}

 \RestyleAlgo{ruled}
\LinesNumbered
\begin{algorithm}
\caption{Signal processing steps}\label{alg:one}
Record electric potential differences \texttt{raw\_ECG} with one reference channel (wrist ECG, Lead I) and multiple head-ECG channels. \\
For data that was recorded as part of Experiment B, perform threshold- and blink-related artifact rejection on \texttt{raw\_ECG}.\\
Bandpass- and notch-filter the reference channel, respectively, using a third-order Butterworth filter with a lower cut-off frequency of $f_{min} =\SI{1}{Hz}$ and an upper cut-off frequency of $f_{max} =\SI{95}{Hz}$, and a second-order IIR-filter with a centre frequency of $f_{c} =\SI{50}{Hz}$ and a band-width of $w=\SI{5}{Hz}$, to give \texttt{filtered\_reference}.\\
Perform R-wave detection on \texttt{filtered\_reference} according to \cite{pan1985real,sedghamiz2014matlab}.\\ 
Bandpass-filter the signals in all channels in \texttt{raw\_ECG} using a third-order Butterworth filter with a lower cut-off frequency of $f_{min}=\SI{1}{Hz}$ and an upper cut-off frequency of $f_{max}=\SI{30}{Hz}$, to give \texttt{filtered\_ECG}.\\
Extract cardiac cycles from \texttt{filtered\_ECG} around the identified R-waves (within a $-\SI{200}{ms}$ to $+\SI{400}{ms}$ window) using the R-wave timings obtained in Step 3 from the wrist ECG.\\ 
Find the median the cardiac rhythms for different lengths of data - ranging from $2-540$ cardiac cycles.\\ 
Calculate four metrics for the quality assessment of the median cardiac cycles from individual channels: (i) correlation between cardiac rhythms in a given channel and Lead I,  (ii) ratios of the amplitude of P-, Q-, S-, and T- waves relative to the R-wave in a given channel, (iii)  timing of P-, Q-, S-, and T-waves relative to the timing of the R-wave in a given channel, and (iv) normalised variance in the channels — the root-mean-square error (RMSE) of the differences between the individual cardiac rhythms and the grand-median cardiac rhythm, divided by the standard deviation of the grand-median cardiac rhythm, in the channel under consideration.
\end{algorithm}

All signal processing was performed in MatLab. The processing steps for the recorded ECG data are described in Algorithm\ref{alg:one}, and consisted of the following procedures. The ECG data from both Experiment A and B were first bandpass-filtered between \SI{0.5}{Hz} and \SI{95}{Hz} using a third-order Butterworth filter and notch-filtered with a second-order IIR-filter with a centre frequency of $f_{c} =\SI{50}{Hz}$ and a band-width of $w=\SI{5}{Hz}$. For data collected in Experiment B (during the driving task), an additional cleaning procedure which removed blink artifacts and large amplitude deflections, such as those caused by motion and jaw-clenching, was applied. Details of the artifact rejection are provided in the supplemental material.
\\

Averaging over multiple consecutive cardiac cycles was conducted in order to extract full cardiac rhythms. In \cite{Wil_earECG}, a mutlimodal in-ear sensor was shown to provide a stand-alone solution for the reliable measurement and identification of cardiac cycles from ear ECG; a collocated MEMS sensor placed beneath the surface of the fabric electrodes was used to detect pressure waves in the ear canal surface associated with the heart beating (e.g., the ballistocardiogram). This served as a reference for the timing of the cardiac cycles in the ear ECG signal detected by the fabric electrodes and, in conjunction with a matched-filtering technique \cite{von2016smart,chanwimalueang2015enabling}, enabled reliable identification of cardiac cycle timing in the low SNR ear ECG signal. Such techniques were not the subject of the present study, therefore timings of the R-peaks that were extracted from the reference channels were used here. However, it is feasible that future developments of the single ear ECG presented in this study will incorporate the described multimodal sensing capabilities. The use of the reference channel in this way is made possible by virtue of the ECG signals from all locations on the body sharing a common electrical source. The R-peak timings were identified using the Pan-Tompkins algorithm implementation in MatLab \cite{sedghamiz2014matlab,pan1985real}. From within the ear ECG signals, segments of \SI{600}{ms} length containing the cardiac rhythms were extracted (\SI{200}{ms} prior to and \SI{400}{ms} after the timing of the R-peak), enabling the identification of all the key components (P-, Q-, R-, S-, and T-waves) in the signal.
\\

Median cardiac rhythms were obtained for different numbers of cardiac cycles \SI{}{N}, ranging from $\SI{}{N}=2$ to $\SI{}{N} = 540$. Since the success of the extraction of cardiac rhythms is dependant on the number of averages taken, and the number of cardiac cycles varies greatly during activities such as driving, different extraction procedures were categorised in terms of the number of cardiac cycles included in the segment, as opposed to the length of time in which the duration took place. During normal heart function (roughly \SI{60}{bpm} - \SI{80}{bpm}), conducting averages over the values of N ranging from $2$ to $540$ produced results that provide an indication of the performance of the ear ECG ranging from recording duration of a few seconds up to roughly 9 minutes. For each value of \SI{}{N} tested, the maximum number of median cardiac rhythms were obtained. For example, for a recording spanning $M = 10$ complete cardiac cycles, an $\SI{}{N}=2$ median cardiac rhythm could be extracted $8$ times ($M - N$). A grand-median cardiac rhythm was extracted from the reference-ECG signal of each subject to serve as a benchmark during analysis of ear ECG cardiac rhythm quality. For rigour, four performance metrics were calculated for each individual ear ECG cardiac rhythm. The four calculated metrics were as follows:

\begin{itemize}
    \item[i)] The Pearson correlation coefficient between the cardiac rhythms from a given channel and a gran-median from the wrist ECG (Lead I);
    \item[ii)] Root mean square (RMS) of the ratios between the R-wave and remaining wave amplitudes in a given channel, bench-marked against equivalent ratios in Lead I; 
    \item[iii)]  Root-mean-square error (RMSE) of the timings of the P-, Q-, S- and T-waves relative to the R-wave in a given channel, relative to timings in lead I. Details of the wave-timing calculation are provided in the supplemental material;
    \item[iv)] RMSE between the grand-median cardiac rhythm of a given channel and all individual cardiac rhythms recorded in that same channel, where the RMSE was normalised by dividing by the standard deviation of the grand-median cardiac rhythm.
\end{itemize}

\begin{figure}[h!]
    \centering
    \includegraphics{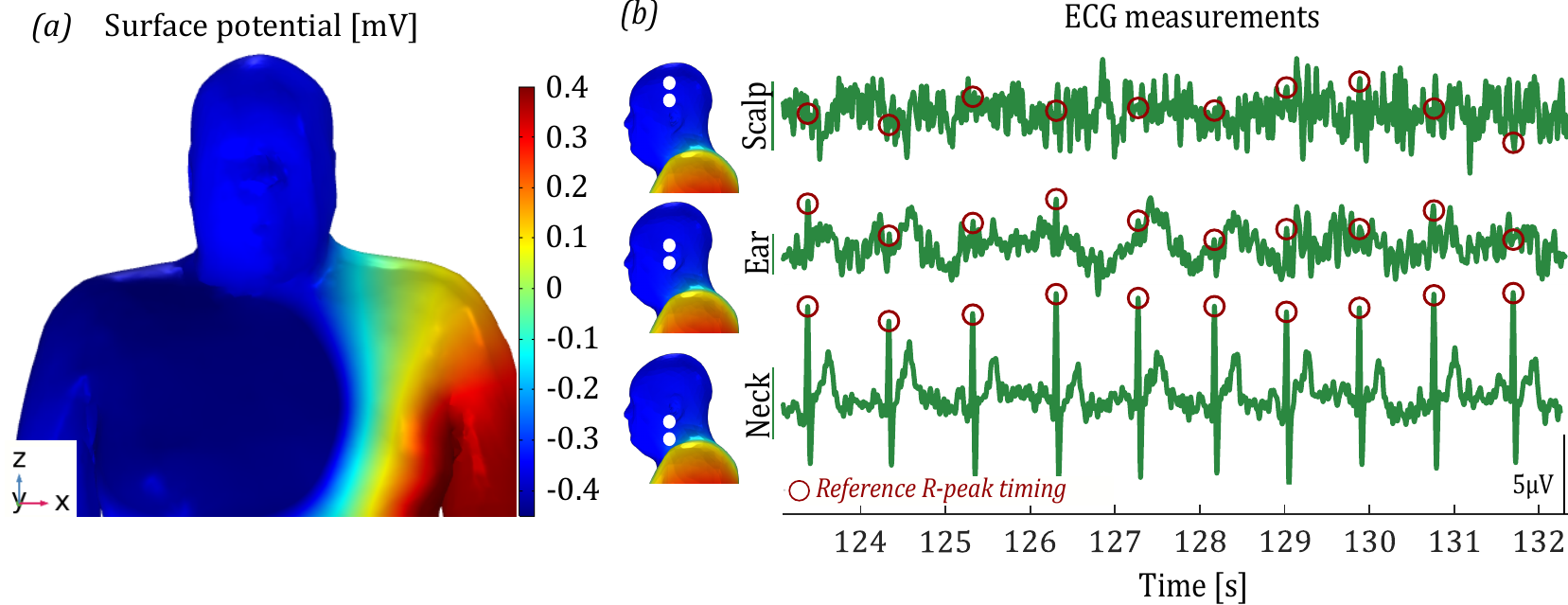}
    \caption{Simulated and recorded cardiac electric potentials. (a) Simulated electric potential on the head surface and upper torso at the time of the R-wave peak (in milliVolts) in the biophysics model. (b) Recorded ECG traces from the scalp, ear, and neck ECG channels from a single subject. The ECG potential at the instance of the R-peaks in the reference (Lead I) channel are indicated (red circle).}
    \label{fig: filtered ecg B}
\end{figure}

\section{Results}
In order to provide rigorous theoretical support to the ear ECG measurements, simulations of ECG propagation throughout the body were conducted. A single ECG cycle was simulated, ranging from the time $t_{1} = -\SI{200}{ms}$ prior to the timing of the maximum of the R-wave, to $t_{2} = \SI{400}{ms}$ after (\SI{600}{ms} in total), at a temporal resolution of $TR = \SI{2}{ms}$. A snap-shot of the so-simulated cardiac cycle at the timing of the maximum of the R-wave is shown in Figure \ref{fig: measurements and model A}a). Notice the large potential difference between the left and right sides of the torso - positions where conventional Lead I electrodes are placed. Observe the reduction in potential difference across the surface of the head (where the considered ear ECG channels are based) relative to the torso. Scalp, ear, and neck ECG cardiac rhythms extracted from virtual sensor positions on a single side of the head (Figure \ref{intro_model_sensors}c)) and are displayed in Figure \ref{fig: correspondence A}a) (blue traces). 

\subsection{Correspondence between simulation and measurement: Experiment A}

\begin{figure}[h!]
    \centering
    \includegraphics{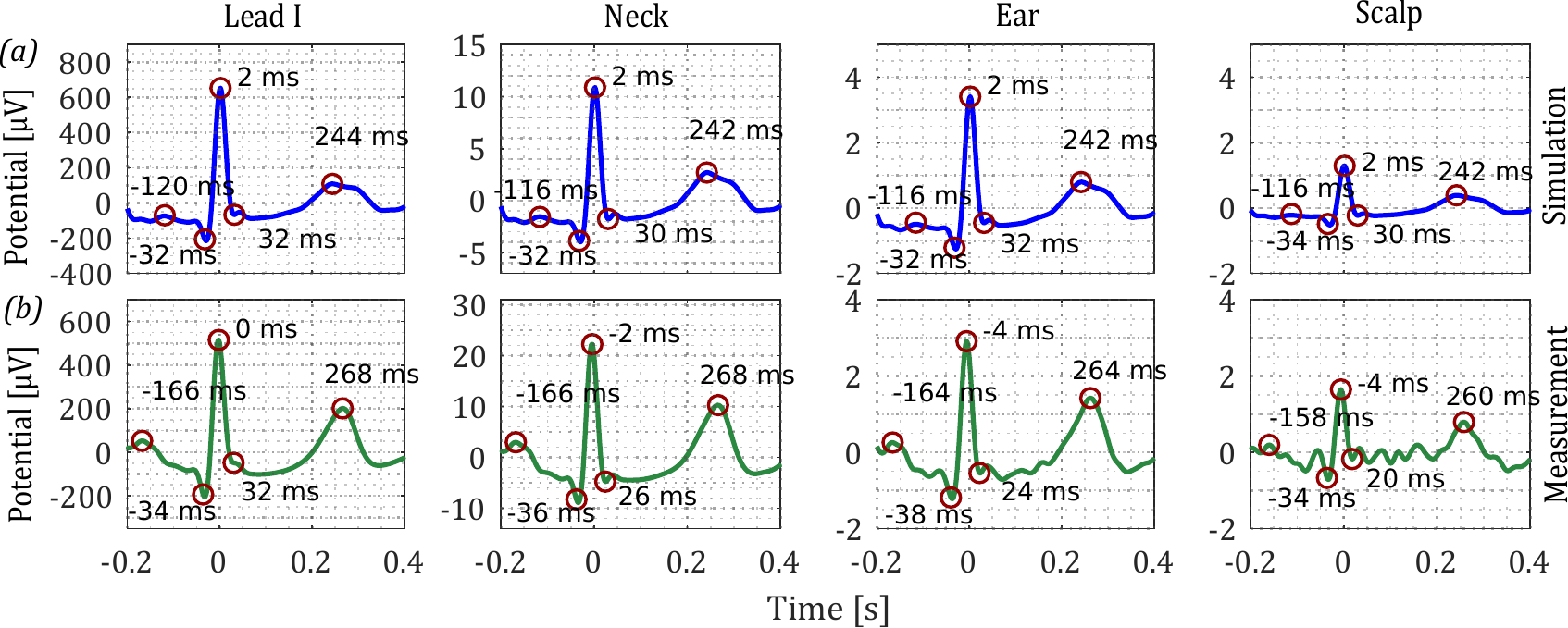}
    \caption{Simulated and measured cardiac rhythms in scalp, ear, and neck ECG. (a) Simulations. (b) Measurements - cardiac rhythms based on averaging over a ten minute recording. ECG potentials as the timings of the P-, Q-, R-, S-, and T-waves (from left to right) are circled. The timing of each eave is indicated in \SI{}{ms}.  }
    \label{fig: correspondence A}
\end{figure}

The narrow neck structure impedes ECG potential as it propagates from the heart to the head surface. It is therefore intuitive that the gradient in the ECG potential will be highest at the base of the neck, where large potential differences begin to occur, and lowest towards the top of the head. Measurements (from Subject 6) in Figure \ref{fig: filtered ecg B} demonstrate such a topography over the neck and head surface; note that the P-,Q-,R-,S-, and T-waves waves in the Lead I cardiac rhythm are identifiable in the neck ECG channel, demonstrating a large potential difference at the position of the neck. On the other hand, in the ear ECG trace, R-peaks are intermittently identifiable, and in the scalp trace, there is no evidence of ECG visible without further processing. Averaging multiple cardiac cycles (e.g., as in Algorithm \ref{alg:one}) enables extraction of the cardiac rhythms from the lower SNR ECG measurements. \\

The model predictions in Figure \ref{fig: correspondence A}a) (blue traces) reveal the characteristic shape and timings of the ECG signal at each location. If there were no other sources (e.g., brain activity and muscle activity) which contributed to the potential difference at these sites, these are the ECG signals that we would expect to obtain. As expected, the amplitude of the ECG on the head decreases the higher up the head the channels are placed, in both the measured and simulated channels. Overall, in both measurements and simulations, the scalp, ear  and neck  signals closely match the Lead I signal. This can be attributed to the fact that the channels under consideration and Lead I form a similar projection plane for the heart vector (Figure \ref{intro_model_sensors}b). Therefore, in this study, the obtained scalp, ear, and neck rhythms were bench-marked against the Lead I equivalent wrist ECG. The measured traces in Figure \ref{fig: correspondence A}b) were obtained by averaging over the entire set of cardiac cycles in the ten minute recording, according to Algorithm \ref{alg:one}. In this way, although high SNR and accurate measurements of cardiac rhythms were extracted, the performance of the channel is not well understood at shorter measurement lengths, when a lower number of cardiac cycles are available.

\begin{figure}[h!]
    \centering
    \includegraphics{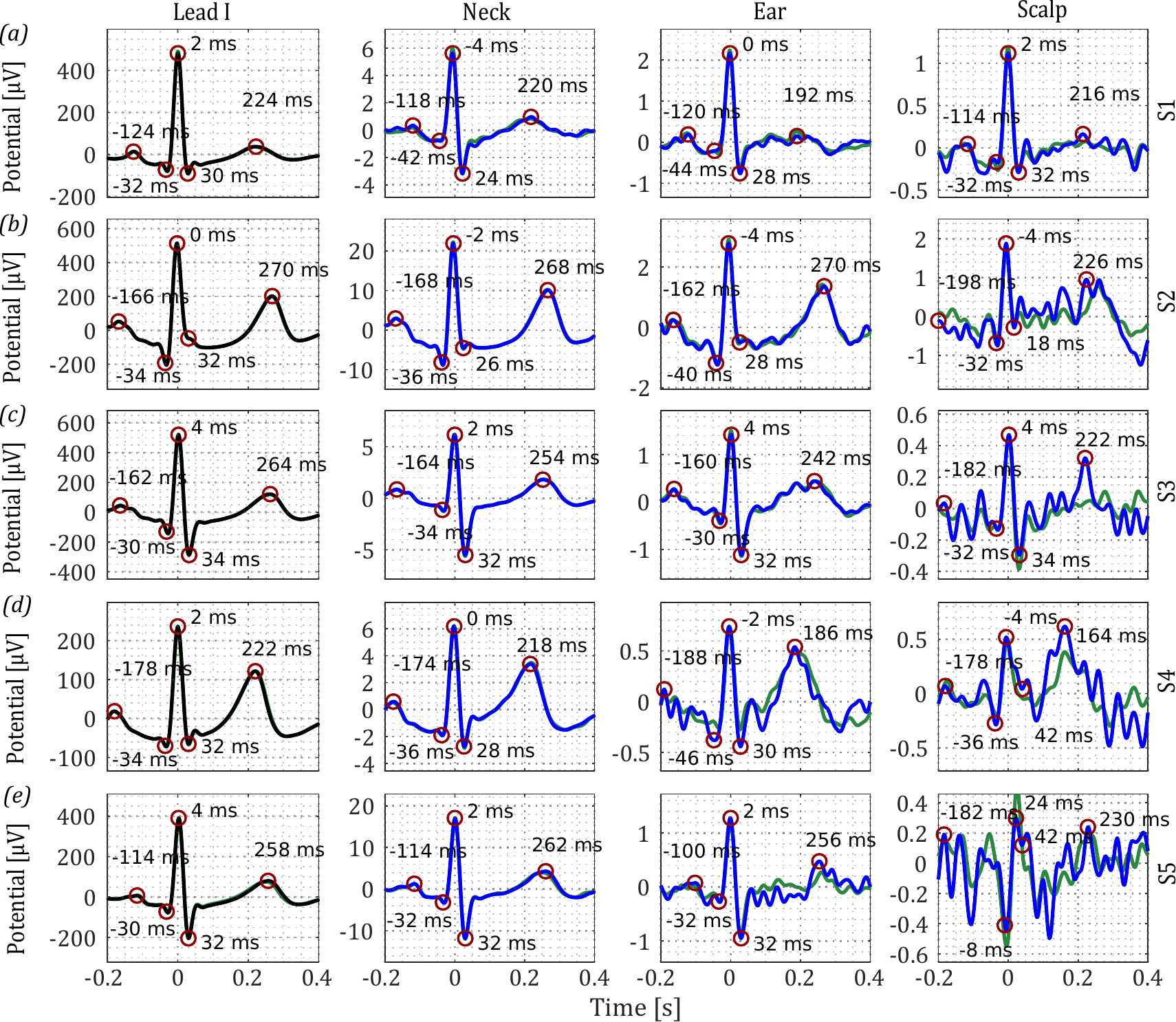}
    \caption{Cardiac rhythms from wrist, scalp, ear, and neck ECG of 5 subjects ((a) - (d)). Reference lead I cardiac cycles are displayed along the first column. ECG acquired after $N=240$ cycles (blue) and $N=540$ cycles (green). ECG potentials at timings of P-, Q-, R-, S-, and T-waves in each channel are circled for the $N=240$ cycle ECG. Timings of the waves are indicated in \SI{}{ms} next to each wave.}
    \label{fig: 5 subjects A}
\end{figure}

\subsection{Ear ECG cardiac rhythms: Experiment A}
\begin{figure}[h!]
    \centering
    \includegraphics{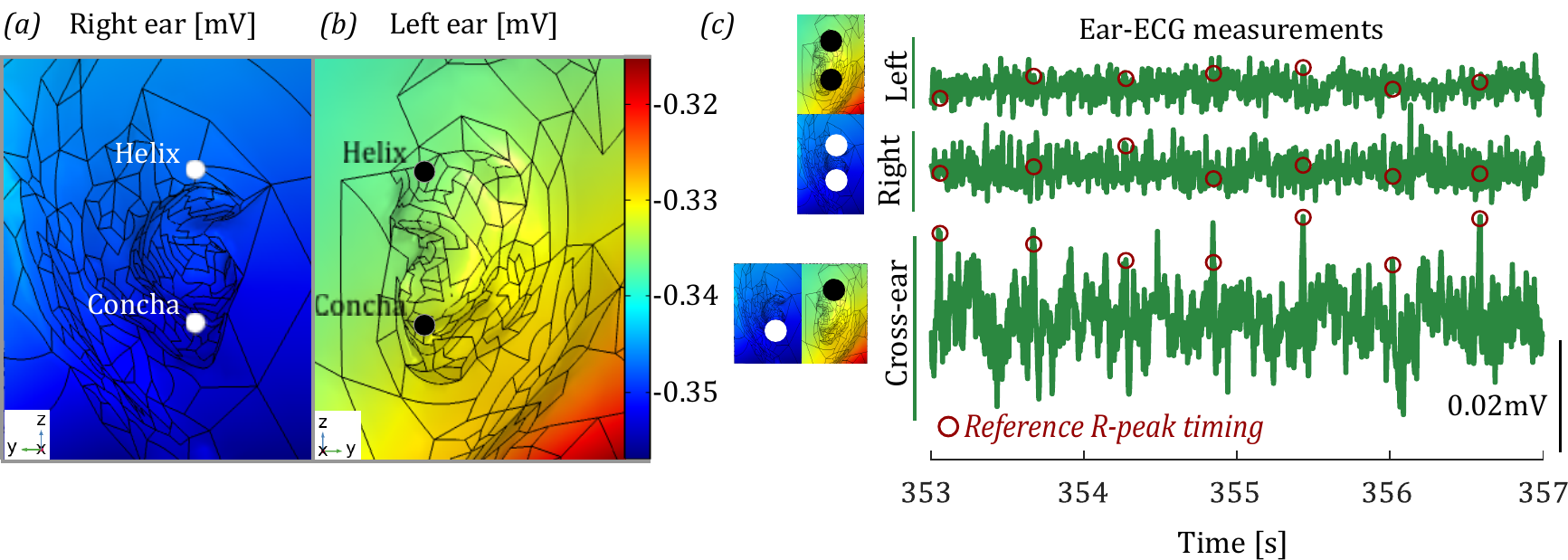}
    \caption{Simulated and recorded cardiac electric potentials. Simulations of (a) left ear and (b) right ear surfaces on the model at the timing of the R-peak. The outline of the mesh of the model is shown in black to improve the clarity of the shape of the ear surface. (c) Recorded ECG traces from left ear, right ear, and cross ear ECG channels from a single subject. ECG potentials in each channel at the timings of the R-peaks in the reference (lead I) channel are circled. }
    \label{fig: measurements and model A}
\end{figure}

The scalp, ear, and neck traces in Figure \ref{fig: measurements and model A} were based on averaging over $540$ cardiac cycles (roughly ten minute recordings), and help establish the potential difference available to the scalp, ear, and neck ECG channels, due to their high SNR. However, in practice, shorter recording lengths are desirable; cardiac rhythms obtained through averaging over the first $240$ recorded cardiac cycles (roughly 4 minutes of data) are shown in Figure \ref{fig: 5 subjects A} for 5 subjects. The cardiac rhythms are plotted in black for the reference-channel and blue for the scalp, ear, and neck ECG channels. While the neck ECG rhythms were consistent across subjects, more variation was observed for the ear and scalp channels. We have included examples of both good and bad  ECG, from Subjects 1-3, and Subject 4-5, respectively, which reflect the level of variation observed across all subjects. Waveforms from the remaining 5 subjects are provided in the supplemental material. The scalp ECG results are considerably worse than the ear  and neck ECG results, which primarily stems from i) poorer electrode skin contact through hair on the scalp, ii) the lower amplitude ECG signal at the location of the scalp relative to the ear and neck, and iii) the large amplitude background noise (EEG and temporal muscle-EMG) on the temporal scalp. Median cardiac cycles from the first $\SI{}{N} = 600$ cycles are also plotted in green. Minimal difference is observed for the high quality neck  and ear ECG channels, however, little change is also observed for the lower quality scalp ECG for Subject 4 and 5; this indicates that a compact device attached on this position on the scalp might not be suitable for ECG recordings.  
\\

\begin{figure}[h]
    \centering
    \includegraphics{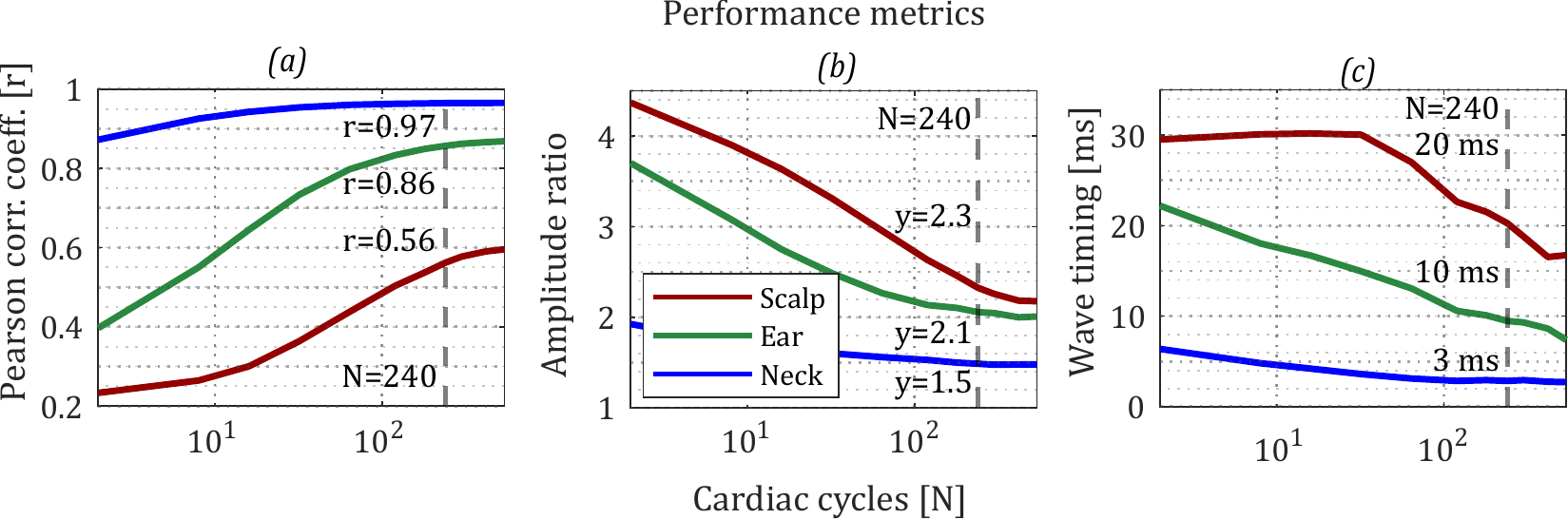}
    \caption{Performance metrics for scalp, ear, and neck ECG after varying levels of averaging (N = cycles). (a) correlation of the cardiac rhythms with the grand-median lead I cardiac rhythm, (b) RMS amplitude ratio between the R-peak and P-,Q-,S-, and T- peaks for a given channel, normalised by the values from lead I (c) RMSE of the timings of the P-,Q-,S-, and T- waves relative to the R-wave between a given channel and the lead I channel. A vertical line indicates the values at $N=240$ cycles. Values for the scalp ECG and neck ECG at $N=240$ cycles are displayed in Table \ref{table metrics}.}
    \label{fig: performance A}
\end{figure}

In instances where urgent care may be required following the diagnosis of an ECG abnormality (such as myocardial infarction following ST elevation), variations in data requirements for the identification of the abnormality are crucial. The reliability of ECG extraction from each channel will be dependant on the number of cycles used to acquire the averaged ECG rhythm, since the averaging process improves the SNR for noise that is uncorrelated with the ECG. Therefore, performance metrics were calculated for averages formed from different numbers of cardiac cycles, ranging from $N=2$ to $N=240$ cardiac cycles (Figure \ref{fig: performance A}). The superiority of the neck ECG is evident, providing high fidelity Lead I ECG after averaging over a low number of cycles, while the scalp ECG performed the worst over the three evaluations. Representative performance metrics after averaging over $N=240$ cycles - a moderate data requirement - are provided in Table \ref{table metrics}. 


\subsection{Correspondence between simulation and measurement: Experiment B}
While the performance of the ear ECG at rest has been demonstrated through the analysis of data collected in Experiment A, the feasibility of recording theses signals in real-world environments must also be established. To that end, measurements during a driving task were conducted on five subjects over the course of one hour. Ear ECG data was collected from both the left and right ears (in a single ear, stand-alone fashion, and from between the ears, as was demonstrated in \cite{Wil_earECG, hammour2019hearables}). 

In order to help provide a better understanding of the ECG potential available at single and cross ear locations, in Figure \ref{fig: filtered ecg B}a), a magnified view of the simulated potential over the ears is provided. The range of potentials spanning the left and right ears (i.e., the maximum available potential from any configuration of ear ECG electrodes) is considerably lower (~$\SI{0.03}{mv}$) than that on the torso ($~\SI{0.8}{mv}$). Moreover, the potential over the surface of a single ear (i.e., the maximum available potential from any configuration of single ear ECG electrodes) is even further reduced to ($~\SI{0.001}{mv}$). Measurements from Subject 4 exemplify the lower amplitude of single ear ECG relative to the cross ear channel; note that the R-peaks are in the bandpass-filtered waveform of the cross ear ECG channel, however, for the single ear ECG, the SNR is too low enable such identification. The reader should note that in the single ear data from Experiment A, R-peaks were identifiable in the bandpass-filtered signal. The difference observed between these waveforms is likely down to the fact that the recordings during Experiment B were conducted in the presence of larger amplitude noise arising due to muscle activity, and after periods of motion of the user, during which the electrode-skin contact can be compromised. 
\\

\begin{figure}[h!]
    \centering
    \includegraphics{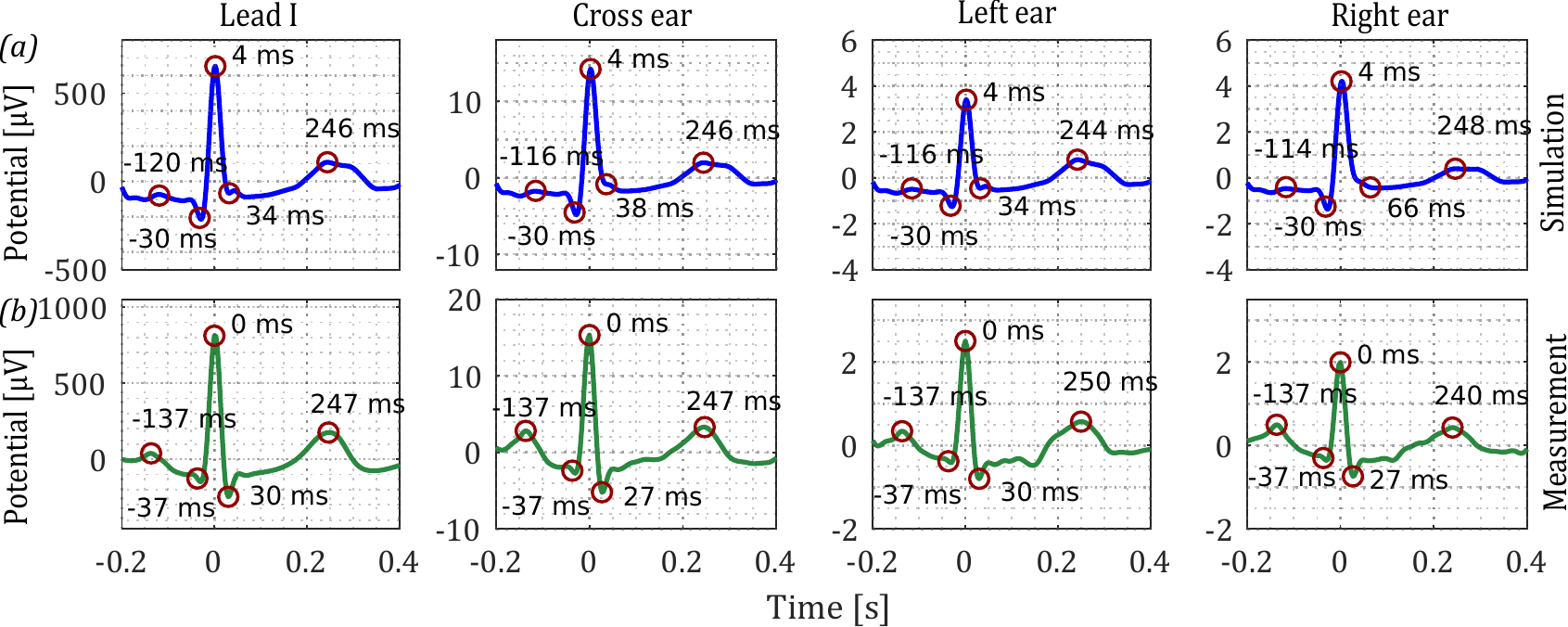}
    \caption{Simulated and measured cardiac rhythms in wrist, left ear, right ear, and cross ear ECG. (a) Simulations. (b) Measurements - cardiac rhythms based on averaging over a sixty minute recording. ECG potentials as the timings of the P-, Q-, R-, S-, and T-waves (from left to right) are circled. The timing of each eave is indicated in \SI{}{ms}.  }
    \label{fig: correspondence B}
\end{figure}

Figure \ref{fig: correspondence B} shows measurements and simulations revealing the characteristic shape and timing of the cardiac rhythms from the single ear and cross ear ECG channels. In both measurements and simulations, two salient features are apparent. The first is the elevated amplitude in the cross ear signal relative to the single ear signals, explained clearly by the surface potentials in Figure \ref{fig: filtered ecg B}a). Next, the good correspondence between the Lead I signal and the cross ear signal is evident (as demonstrated in \cite{Wil_earECG,hammour2019hearables}, as well as the correspondence between the Lead I signal and both single ear signals. However, there is a contradiction between the simulated and measured R-peak amplitudes in the left and right ears, whereby the amplitude of the right ear R-peak is observed to be marginally higher than that in the left ear signal, whereas the measurements across all subjects indicate that the left ear R-peak is the higher (mean = $\SI{0.4}{\mu V}$). This apparent discrepancy should be attributed to the high sensitivity of the model predictions when measuring potential differences across a small distance, whereby small changes in electrode positioning can lead to amplitude changes that are of the same magnitude as the potential difference being measured. Based on the fact that the right ear is slightly further away from the heart than the left ear, one would expect the left ear signal to be slightly higher in amplitude than the right ear signal (as reflected in the measurements), however, further investigation would be needed to establish this difference, for example, through precisely positioned measurements on multiple subjects. 

\subsection{Ear ECG cardiac rhythms: Experiment B}
\begin{figure}[h!]
    \centering
    \includegraphics{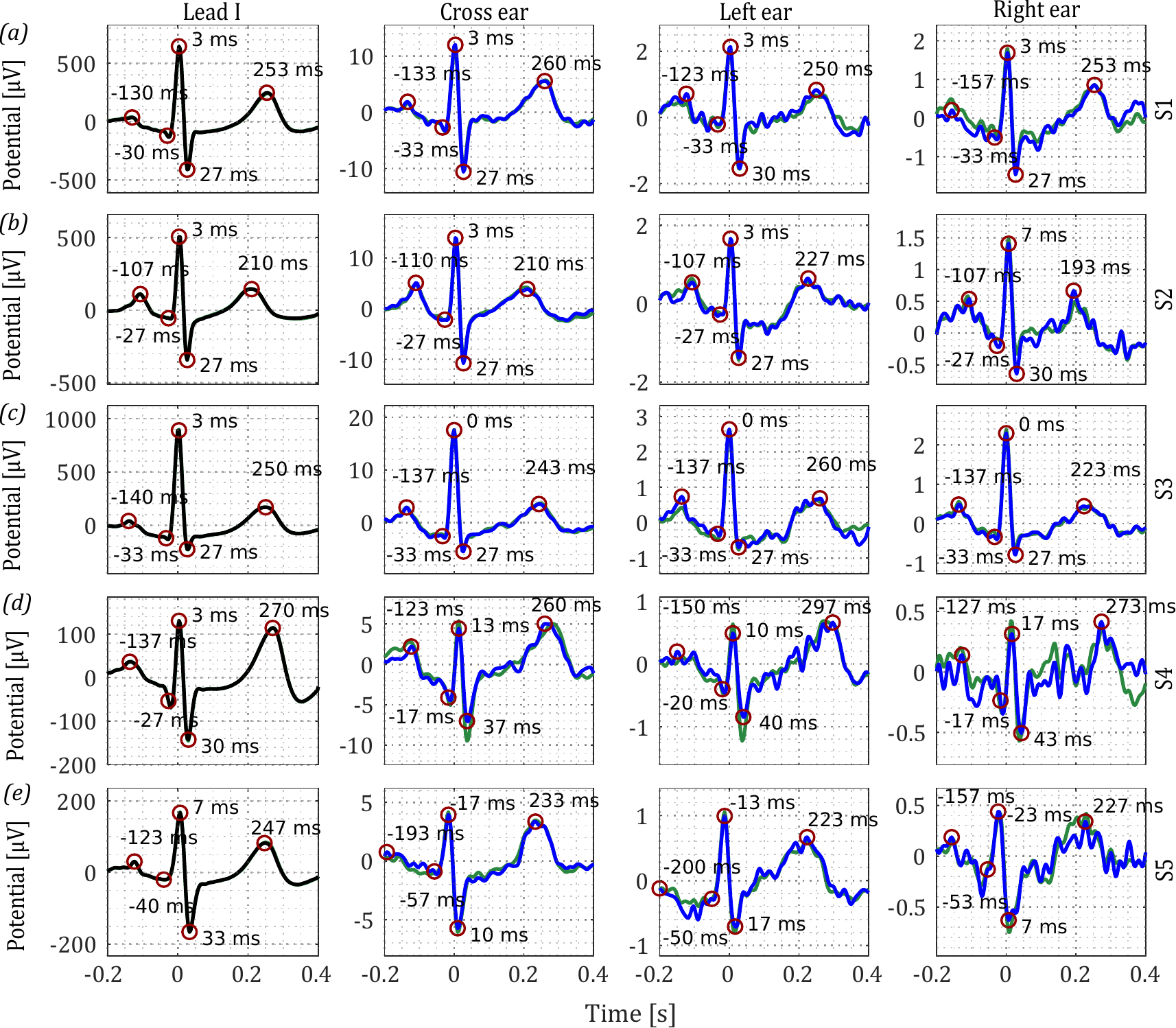}
    \caption{Cardiac rhythms from wrist, left ear, right ear, and cross ear ECG of 5 subjects ((a) - (d)). Reference lead I cardiac cycles are displayed along the first column. ECG acquired after $N=240$ cycles (blue) and $N=600$ cycles (green). ECG potentials at timings of P-, Q-, R-, S-, and T-waves in each channel are circled for the $N=240$ cycle ECG. Timings of the waves are indicated in \SI{}{ms}.}
    \label{fig: 5 subjects B}
\end{figure}

 Cardiac rhythms extracted from $N=240$ cycles for all five subjects are shown in Figure \ref{fig: 5 subjects B}. The displayed cardiac rhythms were taken from the last $240$ cycles recorded during the experiment - such that the impact of real-world recording scenarios (e.g., the presence of muscle movement and physical movement of the user) would be present. Out of the ear ECG channels, the cross ear ECG was more robust, and faithfully retained the Lead I information in each example provided, demonstrating the suitability of this channel to real world recording scenarios. For example, the inter-subject variations in the shape of the Lead I signals are reflected well in the cross ear channel for all five subjects. For the single ear measurements, the Lead I characteristics were still retained, however more distortion in the signal is evident (for Subjects 4 and 5 in particular). Note that, for Subjects 4 and 5, the Lead I amplitude was relatively low compared with Subjects 1-3, indicating that the amplitude of the ECG was low across the whole body for these participants. The implication of this result is that the single ear ECG quality might be highly dependant on the amplitude of electrical activity generated by the heart. For regular ECG, and indeed, even cross-ear ECG, smaller amplitude heart potentials do not severely effect the ECG signal quality,  which provides an explanation for the higher ECG quality. Despite the higher distortion evident in the single ear traces in Figure \ref{fig: 5 subjects B}, for the most part, the multiple waves were still readily discernible. Observe that, as previously discussed, the right ear amplitude was lower than the left ear amplitude. This is also reflected in more distortion visible in the right ear channels. 

Over the course of long-term measurements in real-world scenarios, such as driving, regular movement of the user will induce motion and EMG artifacts in the signal, while also likely compromising the skin-electrode contacts of the channel. In turn, this will all increase the number of cardiac cycles required in order to obtain faithful ECG. Therefore, we have provided an analysis of the performance of the channels with varying data lengths over the course of a one hour driving trial on five subjects.  Performance metrics were calculated on all cardiac cycles recorded during the one hour trial from each subject. Moreover, the performance was evaluated for different levels of averaging (expressed as the number cardiac cycles that were averaged over). Figure \ref{fig: performance A}a) displays the Pearson correlation between the ear ECG cardiac rhythms and Lead I. Correlation results in \cite{Wil_earECG} ($r = 0.96$ for $4$ minutes of cross-ear ECG) from a different cohort of subjects (N = 6) are in good agreement with the results in this study ($r = 0.93$ for $\SI{}{N} = 240$ cycles - or roughly $4$ minutes of data) from five subjects, demonstrating that the performance of the cross-ear ECG can be robust to real-world measurement noise. For single ear ECG, the performance during real world measurements was also stable, with the left single-ear channels performing almost identically in both experiments. For example, the correlation at $N=240$ cycles in experiment A and B, respectively, for the left ear single ear channel were $0.85$ and $0.86$, while the respective timing errors were \SI{10}{ms} and \SI{7}{ms}. Relative to the cross-ear ECG, observe the lower performance of the single ear channel, particularly at smaller values of N. However, at higher values of N, the performance of the single ear channels were similar to the cross-ear channel, indicating that the single ear ECG is also a viable option for cardiac monitoring in real-world scenarios. 

\begin{figure}[h!]
    \centering
    \includegraphics{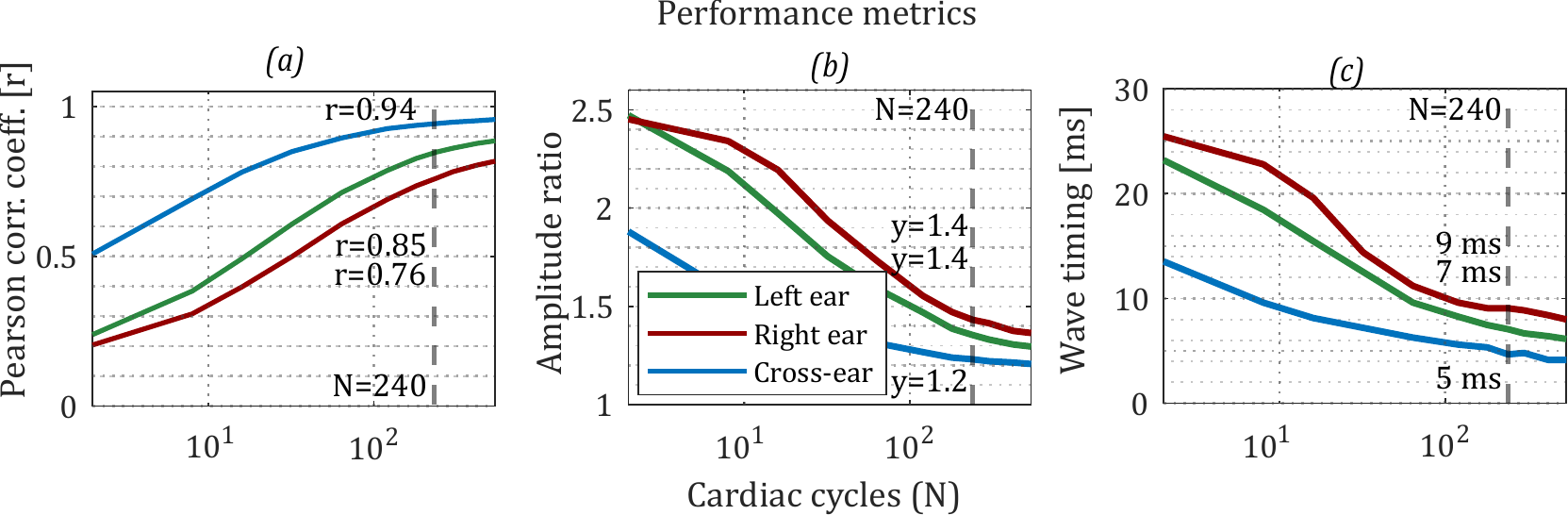}
    \caption{Performance metrics for left ear, right ear, and cross ear ECG after varying levels of averaging (N = cycles). (a) correlation of the cardiac rhythms with the grand-median lead I cardiac rhythm, (b) RMS amplitude ratio between the R-peak and P-,Q-,S-, and T- peaks for a given channel, normalised by the values from lead I (c) RMSE of the timings of the P-,Q-,S-, and T- waves relative to the R-wave between a given channel and the lead I channel. A vertical line indicates the values at $N=240$ cycles, displayed in Table \ref{table metrics}.}
    \label{fig: performance B}
\end{figure}

The observation of a slightly reduced amplitude of the right ear signal relative to that in the left ear is further supported by the performance metrics in Figure \ref{fig: performance B}. However, it is also likely that poorer electrode skin contact would have contributed to the lower performance of the right ear channel.

Table \ref{table metrics} provides performance metrics for the channels under consideration from both Experiment A and B. The values represent the performance of each channel at a value of $N = 240$, which was laso used t obtain the average cardiac ryhtms displayed in Figures \ref{fig: 5 subjects A} and \ref{fig: 5 subjects B}. The performance metrics that are probided in Table \ref{table metrics} represent the mean across all cardiac rhtyhms from within each recording. Performance metrics for the left ear ECG channel were evaluated for data collected in both Experiment A and B. Note that the the performance of the left ear ECG channel (with respect to the timing error and amplitude ratio metrics) was worse in Experiment A relative to Experiment B. This result might seem counter-intuitive, since the data from Experiment A was collected during ideal conditions (with the subject at rest), whereas the data from Experiment B was collected in noise-prone conditions (driving in a simulator environment). However, this discrepancy was possibly caused by the different recording duration over which the performance metrics were evaluated in the two Experiments. For example, at $N=240$ cycles, in recordings from experiment A ($\SI{10}{min.}$ duration), a total of $400$ cardiac rhythms were evaluated, whereas in recordings from experiment B ($\SI{60}{min.}$ duration), roughly $3400$ cardiac rhythms were evaluated.

\begin{table}[h]
\centering
\resizebox{0.8\textwidth}{!}{%
\begin{tabular}{@{}llllllll@{}}
\toprule
          & \multicolumn{2}{l}{i) Correlation} & \multicolumn{2}{l}{ii) Amplitude difference} &  \multicolumn{2}{l}{ii) Time difference}  \\ \midrule     
ECG Channel   & meas.           & sim.          & meas.         & sim.   & meas. (ms)        & sim. (ms)     & iii) var. \\ \midrule
Wrist - rest     & 1                & 1             & 1                & 1           & 0              & 0             & 1 \\
Neck - rest      & 0.97            & 0.99          & 1.5                & 1          & 3             & 1             & 1 \\
cross ear - driving      & 0.94            & 0.99          &1               & 1.1           & 5             & 2            & 3 \\
left ear - rest       & 0.86            & 0.99          & 2.1               &  1.1          & 10          & 2             & 6 \\
left ear - driving       & 0.85            & 0.99          & 1.1               &  1.1          & 7           & 2             & 6 \\
right ear - driving     & 0.76            & 0.99          & 1.3               &  1.2         & 9            &  2             & 9 \\
Scalp - rest     & 0.56            & 0.97          & 2.3                & 1.5          & 20            & 3          & 9
\end{tabular}%
}
\caption{Mean performance for cardiac rhythms ($N=240$ cycles) for scalp, cross ear, left ear, right ear, and neck ECG channels. For the left ear channel, data (i) correlation of the cardiac rhythms with the grand-median lead I cardiac rhythm, (ii) RMS amplitude ratio between the R-peak and P-,Q-,S-, and T- peaks for a given channel, normalise dby the values from lead I iii) RMSE of the timings of the P-,Q-,S-, and T- waves relative to the R-wave between a given channel and the lead I channel, and (iv) normalised variance. The column heading ‘meas.’ denotes results for measured cardiac rhythms, whereas 'sim.' denotes results for the simulated rhythms. Values for left ear, right ear, and cross ear ECG were calculated with data from Experiment B, while values for scalp, and neck ECG were calculated wit data from Experiment A.}
\label{table metrics}
\end{table}

\section{Conclusion}
The ability to monitor ECG will be a key feature in future wearable health systems. A primary position to record physiological signals from is the ear as a result of its stable location relative to the vital organs during everyday activities and its ability to house commonplace accessories such as earbuds. However, the ECG signal that is available over the surface of a single ear had not yet been established. First, the difference in ECG potential on the ear and in surrounding regions of the neck and scalp was investigated. These results can support both existing and prospective wearable ECG platforms that utilise scalp, ear, and neck locations. Measurements of single ear ECG on ten subjects, during an ideal scenario of resting while sitting helped to demonstrate, for the first time, the characteristic timing and shape of the ECG signal available at the single ear location. Further measurements, including both single ear and cross ear ECG, on five subjects during a one hour driving task demonstrated real-world feasibility. Both the cross ear and single ear ECG were demonstrated to be robust to real world environments over prolonged recording periods, providing valuable evidence for the use of such technology in society. Future work will consider the integration of additional cardiac monitoring sensors into the ear worn platform, such as the PPG and BCG, and consider various cardiac conditions.      


\section*{Acknowledgment}
This work was supported by the Racing Foundation grant 285/2018 and the MURI/EPSRC grant EP/P008461.




%
\bibliographystyle{IEEEtran}
\bibliography{references}

%




\end{document}